\documentclass{article}
\usepackage{amssymb}
\usepackage{amsfonts}
\usepackage{amsmath}

\input{tcilatex}

\begin{document}

\title{$d$-dimensional generalization of the point canonical transformation
for a quantum particle with position-dependent mass}
\author{Omar Mustafa$^{1}$ and S.Habib Mazharimousavi$^{2}$ \\
Department of Physics, Eastern Mediterranean University, \\
G Magusa, North Cyprus, Mersin 10,Turkey\\
$^{1}$e-mail: omar.mustafa@emu.edu.tr\\
$^{2}$e-mail: habib.mazhari@emu.edu.tr}
\maketitle

\begin{abstract}
The $d$-dimensional generalization of the point canonical transformation for
a quantum particle endowed with a position-dependent mass in Schr\"{o}dinger
equation is described. Illustrative examples including; the harmonic
oscillator, Coulomb, spiked harmonic, Kratzer, Morse oscillator, P\H{o}%
schl-Teller and Hulth\'{e}n potentials are used as \emph{reference}
potentials to obtain exact energy eigenvalues and eigenfunctions for \emph{%
target} potentials at different position-dependent mass settings.

\end{abstract}

\section{Introduction}

A position-dependent effective mass associated with a quantum mechanical
particle in the Shr\"{o}dinger equation have attracted intense research
activities over the years [1- 14]. It constitutes an interesting and useful
model for the study of many physical problems. In the energy density
functional settings to the many-body problem [1], the non-local term of the
associated potential can be often expressed as a position-dependence of an
appropriate effective mass $M\left( r\right) $. Such an effective mass
concept is used, for example, in the determination of the electronic
properties of the semiconductors [2] and quantum dots [3], in quantum
liquids [4], in $^{3}He$ clusters [5] and metal clusters [6]. Nevertheless,
within the Bohmian approach to quantum theory, the possibility of deriving
the Schr\"{o}dinger equation of particles with position-dependent effective
mass from the Riemannian metric structure is explored and discussed (cf.,
e.g. [7]). Full and partial revivals of a free wave-packet, with
position-dependent effective mass, inside an infinite potential well are
studied and documented [8]. $N$-fold supersymmetry with position-dependent
mass was reported by Tanaka [8], etc.

However, in the study of Hamiltonians for particles endowed with
position-dependent mass, $M\left( r\right) =m_{\circ }m\left( r\right) $,
problems of delicate nature erupt in the process. The momentum operator, for
example, does not commute with $m\left( r\right) $. The choice of the
kinetic energy operator is not unique, hence given rise of a quantum
mechanical problem of long standing known as ordering ambiguity.
Comprehensive details on this issue can be found in the sample of references
in [8].

The above have formed, by large, the manifestos/inspirations of the recent
studies on the one-dimensional Schr\"{o}dinger equation for a particle with
position-dependent effective mass [7-12]. However, conceptual and
fundamental understandings of the quantum physical phenomena may only be
enlightened by the exact solvability of the Schr\"{o}dinger equation. Yet,
such exact solutions form the road-map for improving numerical solutions to
more complicated physical problems.

On the other hand, it is concreted that exactly solvable problems fall
within distinct classes of shape invariant potentials (cf., e.g., [12,13]).
Each of which carries a representation of a dynamical group and can be
mapped into one another by a point canonical transformation (PCT)\ (cf.,
e.g., Alhaidari in [11,12] and Junker in [14]). The Coulomb, the oscillator,
and the s-state Morse problems, for example, belong to the shape invariant
potentials carrying a representation of $so\left( 2,1\right) $ Lie algebra.
In short, a PCT maintains the canonical form of Schr\"{o}dinger equation
invariant.

In the PCT settings, one needs the exact solution of a potential model in a
class of shape invariant potentials to form the so-called \emph{reference
potential}. The \emph{reference potential} \ along with its exact solution
(i.e. eigenvalues and eigenfunctions) is then mapped into the so-called 
\emph{target potential\ ,} hence exact solution for the \emph{target
potential} \ is obtained. Such a recipe is not only bounded to exact
solutions but it is also applicable to the quasi-exact and conditionally
exact ones ( cf, e.g., a sample of references in [15] on the quasi-exact and
conditionally exact solutions), a consensus that should remain beyond doubts
as long as the canonical form remains invariant.

For the sake of completeness, the PCT approach for a quantum particle with a
position-dependent effective mass $M\left( r\right) =m_{\circ }\,m\left(
r\right) $, in Schr\"{o}dinger equation, should be complemented by its $d$%
-dimensional generalization. Where interdimensional degeneracies associated
with the isomorphism between angular momentum $\ell $ and dimensionality $d$
\ are incorporated through the central repulsive/attractive core $\ell
\left( \ell +1\right) /r^{2}\longrightarrow \ell _{d}\left( \ell
_{d}+1\right) /r^{2}$ of the spherically symmetric effective potential $%
V_{eff}\left( r\right) =\ell \left( \ell +1\right) /r^{2}+V\left( r\right) $
(cf, e.g., [16-18] for more details). To the best of our knowledge, the only
attempt was made by Gang [18] on an approximate series solutions of the $d$%
-dimensional position-dependent mass in Schr\"{o}dinger equation.

The forthcoming sections are organized as follows. In section 2 we provide
the $d$-dimensional generalization of the point canonical transformation. We
discuss the consequences of a power-law radial mass in the same section.
Illustrative examples are given in section 3. These examples include; the
harmonic oscillator, the Coulomb, a spiked harmonic, a Kratzer-molecular,
and a Morse oscillator as \emph{reference} potentials. We also give two
illustrative examples on the generalized P\"{o}schl-Teller and Hulth\'{e}n
as \emph{reference} potentials with different position-dependent singular
masses in section 3. Our concluding remarks are given in section 4.\vspace{%
0pt}

\section{PCT $d$-dimensional generalization}

Following the symmetry ordering recipe of the momentum and
position-dependent effective mass ($M(\vec{r})=m_{\circ }\,m\left( \vec{r}%
\right) ,$ and $\alpha =\gamma =0$, and $\beta =-1$ in equation (1.1) of
Tanaka in [8]), the Schr\"{o}dinger Hamiltonian with a potential field $%
V\left( \vec{r}\right) $ would read (in atomic units $\hbar =m_{\circ }=1$)%
\begin{equation}
H=\frac{1}{2}\left( \vec{p}\frac{1}{M\left( \vec{r}\right) }\right) \cdot 
\vec{p}+V\left( \vec{r}\right) =-\frac{\hbar }{2m_{\circ }}\left( \vec{\nabla%
}\frac{1}{m\left( \vec{r}\right) }\right) \cdot \vec{\nabla}+V\left( \vec{r}%
\right) .
\end{equation}%
and assuming the $d$-dimensional spherical symmetric recipe (cf, e.g., Nieto
in [18] for further comprehensive details on this issue), with%
\begin{equation}
\Psi \left( \vec{r}\right) =r^{-\left( d-1\right) /2}R_{n_{r},\ell
_{d}}\left( r\right) Y_{\ell _{d},m_{d}}\left( \theta ,\varphi \right) ,
\end{equation}%
Hamiltonian (1) would result in the following time-independent $d$%
-dimensional radial Schr\"{o}dinger equation%
\begin{equation}
\left\{ \frac{d^{2}}{dr^{2}}-\frac{\ell _{d}\left( \ell _{d}+1\right) }{r^{2}%
}+\frac{m^{\prime }\left( r\right) }{m\left( r\right) }\left( \frac{d-1}{2r}-%
\frac{d}{dr}\right) -2m\left( r\right) \left[ V\left( r\right) -E\right]
\right\} R_{n_{r},\ell }\left( r\right) =0.
\end{equation}%
Where $\ell _{d}=\ell +\left( d-3\right) /2$ for $d\geq 2,$ $\ell $ is the
regular angular momentum quantum number, $n_{r}=0,1,2,\cdots $ is the radial
quantum number, and $m^{\prime }\left( r\right) =dm\left( r\right) /dr.$
Moreover, the $d=1$ can be obtained through $\ell _{d}=-1$ and $\ell _{d}=0$
\thinspace for even and odd parity, $\mathcal{P=}\left( -1\right) ^{\ell
_{d}+1}$, respectively [17]. Nevertheless, the inter-dimensional
degeneracies associated with the isomorphism between angular momentum $\ell $
and dimensionality $d$ \ builds up the ladder of excited states for any
given $n_{r}$ and nonzero $\ell $ from the $\ell =0$ result, with that $%
n_{r} $, by the transcription $d\rightarrow d+2\ell .$ That is, if $%
E_{n_{r},\ell }\left( d\right) $ is the eigenvalue in $d$-dimensions then%
\begin{equation}
E_{n_{r},\ell }\left( 2\right) \equiv E_{n_{r},\ell -1}\left( 4\right)
\equiv \cdots \equiv E_{n_{r},1}\left( 2\ell \right) \equiv
E_{n_{r},0}\left( 2\ell +2\right)
\end{equation}%
for even $d$, and%
\begin{equation}
E_{n_{r},\ell }\left( 3\right) \equiv E_{n_{r},\ell -1}\left( 5\right)
\equiv \cdots \equiv E_{n_{r},1}\left( 2\ell +1\right) \equiv
E_{n_{r},0}\left( 2\ell +3\right)
\end{equation}%
for odd $d$. \ Yet, a unique isomorphism exists between the $S$-wave ($\ell
=0$) energy spectrum in 3D and in 1D (i.e., $E_{n_{r},0}\left( 1\right)
=E_{n_{r},0}\left( 3\right) $). For more details on inter-dimensional
degeneracies the reader may refer to, e.g., [16-18].

A substitution of the form $R\left( r\right) =m\left( r\right) ^{\upsilon
}\,\phi \left( Z\left( r\right) \right) $ in (3) would result in $Z^{\prime
}\left( r\right) =m\left( r\right) ^{1-2\upsilon }$, manifested by the
requirement of a vanishing coefficient of the first-order derivative of $%
\phi \left( Z\left( r\right) \right) $ ( hence a one-dimensional form of Schr%
\"{o}dinger equation is achieved), and $Z^{\prime }\left( r\right)
^{2}=m\left( r\right) $ to avoid position-dependent energies. This, in turn,
mandates $\upsilon =1/4$ and suggests the following point canonical
transformation%
\begin{equation}
q=Z\left( r\right) =\int^{r}\sqrt{m\left( y\right) }dy\text{ }\implies \phi
_{n_{r},\ell _{d}}\left( Z\left( r\right) \right) =m\left( r\right)
^{-1/4}R_{n_{r},\ell _{d}}\left( r\right) .
\end{equation}%
Which in effect implies%
\begin{equation}
\left\{ -\frac{d^{2}}{dq^{2}}+\frac{\ell _{d}\left( \ell _{d}+1\right) }{%
r^{2}m\left( r\right) }+2\left[ V\left( r\right) -U_{d}\left( r\right) -E_{d}%
\right] \right\} \phi _{n_{r},\ell _{d}}\left( q\right) =0,
\end{equation}%
where%
\begin{equation}
U_{d}\left( r\right) =\frac{m^{\prime \prime }\left( r\right) }{8m\left(
r\right) ^{2}}-\frac{7m^{\prime }\left( r\right) ^{2}}{32m\left( r\right)
^{3}}+\frac{m^{\prime }\left( r\right) \left( d-1\right) }{4r\,m\left(
r\right) ^{2}}.
\end{equation}

On the other hand, an exactly solvable (including conditionally-exactly or
quasi-exactly solvable) $d$-dimensional time-independent \thinspace radial
Schr\"{o}dinger \thinspace wave equation (with a constant mass $m_{\circ }$
and cast in $\hbar =m_{\circ }=1$ units)%
\begin{equation}
\left\{ -\frac{d^{2}}{dq^{2}}+\frac{\mathcal{L}_{d}\left( \mathcal{L}%
_{d}+1\right) }{q^{2}}+2\left[ V\left( q\right) -\varepsilon \right]
\right\} \psi _{n_{r},\ell _{d}}\left( q\right) =0
\end{equation}%
would form a \emph{reference} for the exact solvability of the \emph{target}
equation (7). That is, if the exact/conditionally-exact/quasi-exact solution
(analytical/numerical) of (9) is known one can construct the
exact/conditionally-exact/quasi-exact solution of (7) through the relation 
\begin{equation}
\frac{\ell _{d}\left( \ell _{d}+1\right) }{2r^{2}m\left( r\right) }+V\left(
r\right) -U_{d}\left( r\right) -E\iff \frac{\mathcal{L}_{d}\left( \mathcal{L}%
_{d}+1\right) }{2q^{2}}+V\left( q\right) -\varepsilon ,
\end{equation}%
Where $\mathcal{L}_{d}$ is the $d$-dimensional angular momentum quantum
number of the \emph{reference} Schr\"{o}dinger equation.

\subsection{Consequences of a power-law mass $m\left( r\right) =\protect%
\alpha r^{\protect\gamma }$}

With the radial position-dependent mass $m\left( r\right) =\alpha r^{\gamma
} $, the PCT function in (6) implies 
\begin{equation}
Z\left( r\right) =\sqrt{\alpha }\int^{r}y^{\gamma /2}dy=\frac{2\sqrt{\alpha }%
}{\left( \gamma +2\right) }\,r^{\left( \gamma +2\right) /2}\implies \frac{%
\left( \gamma +2\right) }{2}Z\left( r\right) =r\,\sqrt{m\left( r\right) }%
\text{\thinspace }
\end{equation}%
and (8) gives%
\begin{equation}
U_{d}\left( r\right) =-\frac{1}{16}\left( \frac{\gamma \left( 3\gamma
+12-8d\right) }{2r^{2}m\left( r\right) }\right) \equiv -\frac{1}{4}\left( 
\frac{\gamma \left( 3\gamma +12-8d\right) }{2\left( \gamma +2\right)
^{2}Z^{2}\left( r\right) }\right)
\end{equation}%
Relation (10) in effect reads, with $q=Z\left( r\right) $, 
\begin{equation}
\frac{\tilde{\Lambda}\left( \tilde{\Lambda}+1\right) }{2r^{2}m\left(
r\right) }\left( \frac{\gamma }{2}+1\right) ^{2}+V\left( r\right) -E\iff 
\frac{\mathcal{L}_{d}\left( \mathcal{L}_{d}+1\right) }{2q^{2}}+V\left(
q\right) -\varepsilon ,
\end{equation}%
with 
\begin{equation}
\tilde{\Lambda}=-\frac{1}{2}+\left\vert \gamma +2\right\vert ^{-1}\sqrt{%
4\ell _{d}\left( \ell _{d}+1\right) +\left( \gamma -1\right) ^{2}+2\gamma
\left( 3-d\right) }
\end{equation}%
Obviously, Eqs.(11), (13) and (14) suggest that $\gamma =-2$ is not allowed.

\subsection{Remedy at $\protect\gamma =-2$ in a power-law mass $m\left(
r\right) =\protect\alpha r^{\protect\gamma }$}

For the case where $m\left( r\right) =\alpha r^{-2}$ equation (6) implies%
\begin{equation}
q=Z\left( r\right) =\sqrt{\alpha }\int^{r}t^{-1}dt=\sqrt{\alpha }\ln r,
\end{equation}%
and hence%
\begin{eqnarray}
\tilde{U}_{d}\left( \gamma =-2\right) &=&U_{d}\left( r,\gamma =-2\right) -%
\frac{\ell _{d}\left( \ell _{d}+1\right) }{2\alpha }  \notag \\
&=&-\left[ \frac{\left( \ell _{d}+\frac{1}{2}\right) ^{2}+d-1}{2\alpha }%
\right] .
\end{eqnarray}%
Which would only add a constant to the left-hand-side of (10) to yield, with 
$\mathcal{L}_{d}=0$ and/or $\mathcal{L}_{d}=-1$ (i.e., only s-states and/or $%
d=1$ states are available from the right-hand-side of (10) ),%
\begin{equation}
V\left( r\right) -\tilde{U}_{d}\left( \gamma =-2\right) -E\iff V\left(
q\right) -\varepsilon .
\end{equation}

\section{Illustrative examples}

\subsection{$m\left( r\right) =\protect\alpha r^{\protect\gamma }$ with $%
\protect\gamma \neq -2$}

\subsubsection{The harmonic oscillator \emph{reference} potential}

The harmonic oscillator, 
\begin{equation*}
V\left( q\right) =\frac{1}{2}\lambda ^{4}q^{2},
\end{equation*}%
as a \emph{reference} potential, with the exact $d$-dimensional
eigenenergies and wavefunction%
\begin{equation}
\varepsilon _{n_{r},\mathcal{L}_{d}}=\lambda ^{2}\left( 2n_{r}+\mathcal{L}%
_{d}+\frac{3}{2}\right) ,\text{ \thinspace \thinspace }
\end{equation}%
\begin{equation}
\psi _{n_{r},\mathcal{L}_{d}}\left( q\right) =a_{n_{r},\mathcal{L}%
_{d}}\,\left( \lambda q\right) ^{\mathcal{L}_{d}+1}\,\exp \left( -\frac{%
\lambda ^{2}q^{2}}{2}\right) \,\text{\thinspace }L_{n_{r}}^{\mathcal{L}%
_{d}+1/2}\left( \lambda ^{2}q^{2}\right)
\end{equation}%
respectively, would imply a \emph{target} potential 
\begin{equation}
V\left( r\right) =\frac{\omega ^{2}}{2}\alpha r^{\gamma +2};\text{
\thinspace }\omega =\frac{2\lambda ^{2}}{\left( \gamma +2\right) },
\end{equation}%
with corresponding $d$-dimensional eigenenergies and wavefunctions%
\begin{equation}
E_{n_{r},\ell _{d}}=\frac{\left( \gamma +2\right) \omega }{2}\left(
2n_{r}+\Lambda +1\right) ,\text{ \thinspace \thinspace }
\end{equation}%
\begin{equation}
R_{n_{r},\ell _{d}}\left( r\right) =A_{n_{r},\ell _{d}}\,\left( \zeta
r\right) ^{\left( \frac{\gamma }{2}+1\right) \Lambda +\frac{\left( \gamma
+1\right) }{2}}\exp \text{\thinspace }\left( -\frac{\left( \zeta r\right)
^{\gamma +2}}{2}\right) \,L_{n_{r}}^{\Lambda }\left( \left( \zeta r\right)
^{\gamma +2}\right) ,
\end{equation}%
where 
\begin{equation}
\Lambda =\tilde{\Lambda}+1/2=\left\vert \gamma +2\right\vert ^{-1}\sqrt{%
4\ell _{d}\left( \ell _{d}+1\right) +\left( \gamma -1\right) ^{2}+2\gamma
\left( 3-d\right) }
\end{equation}%
and 
\begin{equation}
\zeta =\left[ 2\alpha \omega /\left( \gamma +2\right) \right] ^{1/\left(
\gamma +2\right) }
\end{equation}%
It should be noted that this results, at $d=3$, collapse into Alhaidari's
ones in example 5(a) of his Appendix in [12], where our $\omega $ equals
Alhaidari's $C$.

\subsubsection{The Coulomb \emph{reference} potential}

The Coulomb, 
\begin{equation*}
V\left( q\right) =-A/q,
\end{equation*}%
as a \emph{reference} potential, with the exact $d$-dimensional
eigenenergies and wavefunction%
\begin{equation}
\varepsilon _{n_{r},\mathcal{L}_{d}}=-\frac{\text{\thinspace }\lambda
_{n_{r},\mathcal{L}_{d}}^{2}}{8};\text{ \thinspace }\lambda _{n_{r},\mathcal{%
L}_{d}}=\frac{2A}{\left( n_{r}+\mathcal{L}_{d}+1\right) },
\end{equation}%
\begin{equation}
\text{\thinspace }\psi _{n_{r},\mathcal{L}_{d}}\left( q\right) =N_{n_{r},%
\mathcal{L}_{d}}\,\,q^{\mathcal{L}_{d}+1}\,\exp \left( -\frac{\lambda
_{n_{r},\mathcal{L}_{d}}\,q}{2}\right) \,\text{\thinspace }L_{n_{r}}^{2%
\mathcal{L}_{d}+1}\left( \lambda _{n_{r},\mathcal{L}_{d}}\,q\right)
\end{equation}%
respectively, would imply a \emph{target} potential%
\begin{equation}
V(r)=-\frac{C}{2\sqrt{\alpha }}\,r^{-1-\gamma /2};\text{ \thinspace }%
C=A\left( \gamma +2\right) ,
\end{equation}%
with corresponding $d$-dimensional eigenenergies and wavefunctions%
\begin{equation}
E_{n_{r},\ell _{d}}=-\frac{C^{2}/2}{\left( \gamma +2\right) ^{2}}\frac{1}{%
\left( n_{r}+\Lambda +1/2\right) ^{2}}
\end{equation}%
\begin{equation}
R_{n_{r},\ell _{d}}\left( r\right) =A_{n_{r},\ell _{d}}\,\left( \tilde{\zeta}%
\,r\right) ^{\left( \frac{\gamma }{2}+1\right) \Lambda +\frac{\left( \gamma
+1\right) }{2}\,}\exp \left( -\frac{\left( \tilde{\zeta}\,r\right) ^{\frac{%
\gamma }{2}+1}}{2}\right) L_{n_{r}}^{2\Lambda }\left( \left( \tilde{\zeta}%
\,r\right) ^{\frac{\gamma }{2}+1}\right)
\end{equation}%
where%
\begin{equation}
\tilde{\zeta}=\tilde{\zeta}\left( n_{r},\ell _{d}\right) \,=\left[ \frac{4C%
\sqrt{\alpha }}{\left( \gamma +2\right) ^{2}}\left( n_{r}+\Lambda
+1/2\right) ^{-1}\right] ^{1/\left( \frac{\gamma }{2}+1\right) }
\end{equation}%
It should be noted that our results (27)-(30), at $d=3$, collapse into
Alhaidari's ones reported in example 5(b) of his Appendix in [12], as his
second solution of his Eq. (3.3) using the \emph{reference} potential 3D
harmonic oscillator. It seems that, in Alhaidari's second solution proposal
of his Eq.(3.3) there is an implicit latent additional change of variables
of a Liouvillean nature (cf., e.g., [24-26]) that led to a Coulomb-harmonic
oscillator correspondence ( the reader may wish to investigate this issue
following, e.g.,Znojil and L\'{e}vai [26]). The proof of which is beyond our
current proposal.

\subsubsection{A spiked harmonic oscillator \emph{reference} potential}

A spiked harmonic oscillator ( or a Gold'man and Krivchenkov model), 
\begin{equation*}
V(q)=\lambda ^{4}q^{2}/2+\beta q^{-2}/2,
\end{equation*}%
as a \emph{reference} potential (cf,e.g.,[20]), with the exact $d-$%
dimensional eigenenergies and wavefunction%
\begin{equation}
\varepsilon _{n_{r},\mathcal{L}_{d}}=\lambda ^{2}\left( 2n_{r}+\mathcal{%
\tilde{L}}_{d}+\frac{3}{2}\right) ,\text{ \thinspace \thinspace }\mathcal{%
\tilde{L}}_{d}=-\frac{1}{2}+\sqrt{\left( \mathcal{L}_{d}+\frac{1}{2}\right)
^{2}+\beta }.
\end{equation}%
\begin{equation}
\psi _{n_{r},\mathcal{L}_{d}}\left( q\right) =a_{n_{r},\mathcal{L}%
_{d}}\,\left( \lambda q\right) ^{\mathcal{\tilde{L}}_{d}+1}\,\exp \left( -%
\frac{\lambda ^{2}q^{2}}{2}\right) \,\text{\thinspace }L_{n_{r}}^{\mathcal{%
\tilde{L}}_{d}+1/2}\left( \lambda ^{2}q^{2}\right)
\end{equation}%
would lead to a \emph{target} potential%
\begin{equation}
V\left( r\right) =\frac{\omega ^{2}}{2}\alpha r^{\gamma +2}+\frac{\tilde{%
\beta}}{2\alpha }r^{-\gamma -2};\text{ \thinspace \thinspace }\omega =\frac{%
2\lambda ^{2}}{\left( \gamma +2\right) },\,\text{\thinspace }\tilde{\beta}=%
\frac{\beta \left( \gamma +2\right) ^{2}}{4}
\end{equation}%
with corresponding $d$-dimensional eigenenergies and wavefunctions%
\begin{equation}
E_{n_{r},\ell _{d}}=\frac{\left( \gamma +2\right) \omega }{2}\left(
2n_{r}+\delta +1\right) \text{ ; }\delta =\sqrt{\Lambda ^{2}+\beta }\text{
\thinspace }
\end{equation}%
\begin{equation}
\text{\thinspace }R_{n_{r},\ell _{d}}\left( r\right) =N_{n_{r},\ell
_{d}}\,\left( \zeta r\right) ^{\left( \frac{\gamma }{2}+1\right) \delta
+\left( \gamma +1\right) /2}\exp \left( -\frac{\left( \zeta r\right)
^{\gamma +2}}{2}\right) L_{n_{r}}^{\delta }\left( \left( \zeta r\right)
^{\gamma +2}\right) ,
\end{equation}%
It should be noted that Eq.(31) reduces to Eq.(12) of Yu and Dong [10] when $%
d=1$, $\mathcal{L}_{d}=0,-1,$ and $\gamma =-3$ to read%
\begin{equation*}
E_{n_{r},0}=\lambda ^{2}\left( 2n_{r}+3\right) =\sqrt{2A}\left(
2n_{r}+3\right)
\end{equation*}%
where our $\lambda ^{2}=\sqrt{2A}$, $A$ is defined by Yu and Dong [10] as $%
A=\xi \tau ^{2}/4$, and our $\beta /2=15/8$. One can also show that $u$ in
Yu and Dong is equal to $\left[ \lambda ^{2}q^{2}\right] $, and hence the
corresponding wave function, for the $d=1$ case,%
\begin{equation*}
\psi _{n_{r},\mathcal{L}_{d}}\left( q\right) =N_{n_{r}}\,\left( \sqrt{u}%
\right) ^{\Lambda +1/2}\,\exp \left( -\frac{u}{2}\right) \,\text{\thinspace }%
L_{n_{r}}^{\Lambda }\left( u\right) ,
\end{equation*}%
with $\Lambda =2$, is exactly the same as that in Eq.(11) of Yu and Dong in
[10]. Moreover, our $V\left( r\right) $ in (33) is the same as Eq. (4b) Yu
and Dongs, of course with the proper amendments.

\subsubsection{A Kratzer's-type \emph{reference} potential}

A Kratzer's-type molecular potential (cf, e.g., Fl\"{u}gge in [21]), 
\begin{equation*}
V(q)=-A/q+\beta q^{-2}/2,
\end{equation*}%
as a \emph{reference} potential with the exact $d-$dimensional eigenenergies
and wavefunction 
\begin{equation}
\varepsilon _{n_{r},\mathcal{L}_{d}}=-\frac{\text{\thinspace }\tilde{\lambda}%
_{n_{r},\mathcal{L}_{d}}^{2}}{8};\text{ \thinspace }\tilde{\lambda}_{n_{r},%
\mathcal{L}_{d}}=\frac{2A}{\left( n_{r}+\mathcal{\tilde{L}}_{d}+1\right) },
\end{equation}%
\begin{equation}
\text{\thinspace }\psi _{n_{r},\mathcal{L}_{d}}\left( q\right) =N_{n_{r},%
\mathcal{L}_{d}}\,\,q^{\mathcal{\tilde{L}}_{d}+1}\,\exp \left( -\frac{\tilde{%
\lambda}_{n_{r},\mathcal{L}_{d}}\,q}{2}\right) \,\text{\thinspace }%
L_{n_{r}}^{2\mathcal{\tilde{L}}_{d}+1}\left( \tilde{\lambda}_{n_{r},\mathcal{%
L}_{d}}\,q\right)
\end{equation}%
respectively, would imply a set of \emph{target} potentials%
\begin{equation}
V(r)=-\frac{C}{2\sqrt{\alpha }}\,r^{-1-\gamma /2}+\frac{\tilde{\beta}}{%
2\alpha }r^{-\gamma -2}\text{ \thinspace },
\end{equation}%
where $C=A\left( \gamma +2\right) ,$ \thinspace $\tilde{\beta}=\beta \left(
\gamma +2\right) ^{2}/4$, and corresponding $d$-dimensional eigenenergies
and wavefunctions%
\begin{equation}
E_{n_{r},\ell _{d}}=-\frac{C^{2}/2}{\left( \gamma +2\right) ^{2}}\frac{1}{%
\left( n_{r}+\delta +1/2\right) ^{2}};\text{ \thinspace \thinspace
\thinspace }\delta =\sqrt{\Lambda ^{2}+\beta }
\end{equation}%
\begin{equation}
R_{n_{r},\ell _{d}}\left( r\right) =A_{n_{r},\ell _{d}}\,\left( \eta
\,r\right) ^{\left( \frac{\gamma }{2}+1\right) \delta +\frac{\left( \gamma
+1\right) }{2}}\,\exp \left( -\frac{\left( \eta \,r\right) ^{\frac{\gamma }{2%
}+1}}{2}\right) L_{n_{r}}^{2\delta }\left( \left( \eta \,r\right) ^{\gamma
/2+1}\right)
\end{equation}%
where%
\begin{equation}
\eta =\eta \left( n_{r},\ell _{d}\right) \,=\left[ \frac{4C\sqrt{\alpha }}{%
\left( \gamma +2\right) ^{2}}\left( n_{r}+\delta +1/2\right) ^{-1}\right]
^{1/\left( \frac{\gamma }{2}+1\right) }
\end{equation}

\subsection{$m\left( r\right) =\protect\alpha r^{\protect\gamma }$ with $%
\protect\gamma =-2$}

\subsubsection{The spiked harmonic oscillator \emph{reference} potential}

The spiked harmonic oscillator ( or a Gold'man and Krivchenkov model), 
\begin{equation*}
V(q)=\lambda ^{4}q^{2}/2+\beta q^{-2}/2,
\end{equation*}%
as a \emph{reference} potential (cf,e.g.,[20]), with the exact $d-$%
dimensional s-states' eigenenergies and wavefunctions%
\begin{equation}
\varepsilon _{n_{r},0}=\lambda ^{2}\left( 2n_{r}+k_{d}+\frac{3}{2}\right) ,%
\text{ \thinspace }k_{d}=-\frac{1}{2}+\sqrt{\left( \frac{1}{2}\right)
^{2}+\beta }.
\end{equation}%
\begin{equation}
\psi _{n_{r},0}\left( q\right) =a_{n_{r},0}\,\left( \lambda q\right)
^{k_{d}+1}\,e^{-\lambda ^{2}q^{2}/2}\,\text{\thinspace }%
L_{n_{r}}^{k_{d}+1/2}\left( \lambda ^{2}q^{2}\right)
\end{equation}%
would lead to a \emph{target} potential%
\begin{equation}
V\left( r\right) =\frac{1}{2\alpha }\left( \ln r\right) ^{2}+\frac{C^{2}}{2}%
\left( \ln r\right) ^{-2};\text{ \thinspace \thinspace }\alpha =\lambda
^{-2},\,\text{\thinspace }C^{2}=\frac{\beta }{\alpha }
\end{equation}%
with a corresponding $d$-dimensional eigenenergies and wavefunctions%
\begin{equation}
E_{n_{r},\ell _{d}}=\frac{1}{\alpha }\left( 2n_{r}+\Omega +\frac{\left( \ell
_{d}+\frac{1}{2}\right) ^{2}+d+1}{2}\right) ,\,\Omega =\text{\thinspace }%
2^{-1}\sqrt{1+4\alpha \text{\thinspace }C^{2}}\text{ \thinspace }
\end{equation}%
\begin{equation}
\text{\thinspace }R_{n_{r}}\left( r\right) =B_{n_{r}}\,\frac{1}{\sqrt{r}\,}%
\left( \ln r\right) ^{\Omega +1/2}\,\exp \text{\thinspace }\left[ -\frac{%
\left( \ln r\right) ^{2}}{2}\right] \,L_{n_{r}}^{\Omega }\left( \left( \ln
r\right) ^{2}\right) ,
\end{equation}%
It should be reported here that equations (45) and (46) reduce to the
results obtained by Alhaidari (see example 5 in the Appendix of [12]) \ for $%
\ell _{d}=0$ and $d=3$, However, it is worthy to mention that the $\ell _{d}$%
-dependence of the energy eigenvalues of the \emph{target} potential are
manifested by the consideration of the constant term in (16) of our proposal.

\subsubsection{A Kratzer's-type molecular \emph{reference} potential}

A Kratzer's-type molecular potential, 
\begin{equation*}
V(q)=-A/q+\beta q^{-2}/2,
\end{equation*}%
as a \emph{reference} potential with the exact $d$-dimensional s-states'
eigenenergies and wavefunctions 
\begin{equation}
\varepsilon _{n_{r},0}=-\frac{\text{\thinspace }\tilde{\lambda}_{n_{r},0}^{2}%
}{8};\text{ \thinspace }\tilde{\lambda}_{n_{r},0}=\frac{2A}{\left(
n_{r}+k_{d}+1\right) },
\end{equation}%
\begin{equation}
\text{\thinspace }\psi _{n_{r},0}\left( q\right)
=N_{n_{r},0}\,\,q^{k_{d}+1}\,\exp \left( -\frac{\tilde{\lambda}_{n_{r},0}\,q%
}{2}\right) \,\text{\thinspace }L_{n_{r}}^{2k_{d}+1}\left( \tilde{\lambda}%
_{n_{r},0}\,q\right)
\end{equation}%
would lead to the \emph{target} potential%
\begin{equation}
V\left( r\right) =-\frac{A}{\sqrt{\alpha }\ln r}+\frac{\beta }{2\alpha
\left( \ln r\right) ^{2}}
\end{equation}%
with corresponding $d$-dimensional eigenenergies and wavefunctions%
\begin{equation}
E_{n_{r},\ell _{d}}=\frac{\left( \ell _{d}+\frac{1}{2}\right) ^{2}+d-1}{%
2\alpha }-\frac{\text{\thinspace }\tilde{\lambda}_{n_{r},0}^{2}}{8},
\end{equation}%
\begin{gather}
R_{n_{r}}\left( r\right) =N_{n_{r}}\,\frac{1}{\sqrt{r}}\left( \,\ln r\right)
^{k_{d}+1}\,\exp \text{\thinspace }\left[ -\frac{\tilde{\lambda}_{n_{r},0}%
\sqrt{\alpha }\ln r}{2}\right] \,  \notag \\
\times L_{n_{r}}^{2k_{d}+1}\left( \lambda _{n_{r},0}\sqrt{\alpha }\ln
r\right) .
\end{gather}

\subsubsection{A Morse-oscillator \emph{reference} potential}

A Morse-oscillator potential of the form 
\begin{equation*}
V\left( q\right) =Ae^{-2aq}-Be^{-aq}\text{ };\text{ \ }B=2A,
\end{equation*}%
as a \emph{reference} potential (cf, e.g., [19]) with $\mathcal{L}_{d}=0$
and/or $\mathcal{L}_{d}=-1,$with the exact $s$-states $d$-dimensional
eigenenergies wavefunctions%
\begin{equation}
\varepsilon _{n_{r}}=-A\left[ 1-\sqrt{\frac{1}{2A\alpha }}\left( n_{r}+\frac{%
1}{2}\right) \right] ^{2}\text{ };\text{ \ }a=1/\sqrt{\alpha }
\end{equation}%
\begin{equation}
\psi _{n_{r}}\left( q\right) =N_{n_{r}}\,u^{s}\,e^{-u/2}\,F\left(
-n_{r},2s+1,u\right) ;
\end{equation}%
\begin{equation*}
\text{ \thinspace }u=\sqrt{8\alpha A}e^{-aq},\text{ }s=\sqrt{-2\alpha
\varepsilon _{n_{r}}}.
\end{equation*}%
would lead to a \emph{target} potential%
\begin{equation}
V(r)=-A\left( \frac{1}{r^{2}}-\frac{2}{r}\right)
\end{equation}%
with corresponding $d$-dimensional eigenenergies and wavefunctions%
\begin{equation}
E_{n_{r},\ell _{d}}=\frac{\left( \ell _{d}+\frac{1}{2}\right) ^{2}+d-1}{%
2\alpha }-A\left[ 1-\frac{1}{\sqrt{2A\alpha }}\left( n_{r}+\frac{1}{2}%
\right) \right] ^{2}
\end{equation}%
\begin{equation}
\text{\thinspace }R_{n_{r}}\left( r\right) =\tilde{N}_{n_{r}}\,\left( \frac{1%
}{r}\right) ^{s+1/2}e^{-u/2}F\left( -n_{r},2s+1,u\right) .
\end{equation}%
Where $u=\sqrt{8\alpha A}/r$. It should reported here that when $d=1$ and $%
\ell _{d}=0,-1$ Eq.(55) reduces to Eq.s (14) and (!5) in [10].

\subsection{Two example on $m\left( r\right) \neq \protect\alpha r^{\protect%
\gamma }$}

\subsubsection{ A Periodic Generalized P\"{o}schl-Teller \emph{reference}
potential and $m\left( r\right) =\protect\alpha /4r\left( 1+r\right) ^{2}$ $%
\implies $ $q\left( r\right) =\protect\sqrt{\protect\alpha }\arctan \protect%
\sqrt{r}$}

A Generalized P\"{o}schl-Teller potential [21] of the form 
\begin{equation}
V\left( q\right) =\frac{\zeta ^{2}}{2}\left[ \frac{\tau \left( \tau
-1\right) }{\cos ^{2}\zeta q}+\frac{\varkappa \left( \varkappa -1\right) }{%
\sin ^{2}\zeta q}\right]
\end{equation}%
as a \emph{reference} potential with the exact $d$-dimensional $s$-states'
eigenvalues and eigenfunctions%
\begin{equation}
\varepsilon _{n_{r},0}=\frac{1}{2}\zeta ^{2}\left[ \varkappa +\tau +2n_{r}%
\right] ^{2}
\end{equation}%
\begin{equation}
\psi _{n_{r}}\left( q\right) =C_{n_{r},\varkappa ,\tau }\left( \sin \zeta
q\right) ^{\varkappa }\,\left( \cos \zeta q\right) ^{\tau }\,_{2}F_{1}\left(
-n_{r},\varkappa +\tau +n_{r},\varkappa +1/2;\sin ^{2}\zeta q\right)
\end{equation}%
would lead to a \emph{target} potential%
\begin{equation}
V\left( r\right) =V_{1}\left( 1+r^{2}\right) +V_{2}\left( 1+\frac{1}{r^{2}}%
\right)
\end{equation}%
with $d$-dimensional $s$-states' eigenenergies and eigenfunctions%
\begin{equation}
E_{n_{r},\ell _{d}}=\frac{1}{2}\zeta ^{2}\left[ \varkappa +\tau +2n_{r}%
\right] ^{2}-\left( \eta _{1}+\eta _{2}+\eta _{3}\right)
\end{equation}%
\begin{align}
\text{\thinspace }R_{n_{r}}\left( r\right) & =C_{n_{r},\varkappa ,\tau }%
\left[ \frac{\alpha }{4r\left( 1+r\right) ^{2}}\right] ^{\frac{1}{4}}\left[ 
\frac{r}{1+r}\right] ^{\frac{\varkappa }{2}}  \notag \\
& \times \left[ \frac{1}{1+r}\right] ^{\tau }\,_{2}F_{1}\left(
-n_{r},\varkappa +\tau +n_{r},\varkappa +1/2;\frac{r}{1+r}\right) .
\end{align}%
Where 
\begin{gather}
\zeta =1/\sqrt{\alpha },\eta _{1}=\left( 24d-9\right) /8\alpha ,\eta
_{2}=\left( 8d-9\right) /8\alpha ,  \notag \\
\eta _{3}=-\left( 32d-22\right) /8\alpha
\end{gather}%
and 
\begin{eqnarray}
V_{1} &=&\frac{\varkappa \left( \varkappa -1\right) -\left( 2\alpha \eta
_{1}+4\ell _{d}\left( \ell _{d}+1\right) \right) }{2\alpha }  \notag \\
V_{2} &=&\frac{\tau \left( \tau -1\right) -\left( 2\alpha \eta _{2}+4\ell
_{d}\left( \ell _{d}+1\right) \right) }{2\alpha }
\end{eqnarray}

\subsubsection{A Generalized Hulth\'{e}n \emph{reference} potential and $%
m(r)=1/\protect\alpha ^{2}(r+1)^{2}$ $\implies $ $q\left( r\right) =\protect%
\alpha ^{-1}\ln \left( r+1\right) $}

A Generalized Hulth\'{e}n potential [22] of the form%
\begin{equation}
V\left( q\right) =-\frac{\alpha e^{-\alpha q}}{1-e^{-\alpha q}}
\end{equation}%
as a \emph{reference} potential with the exact $d$-dimensional eigenvalues
and eigenfunctions (for the $s$ -states)%
\begin{equation}
\varepsilon _{n_{r},0}=\frac{\alpha ^{2}}{2}Q_{n_{r}}^{2}
\end{equation}%
\begin{equation}
\psi _{n_{r}}\left( q\right) =C_{n_{r},0}e^{-Q_{n_{r}}\alpha
q}\dsum\limits_{\nu =1}^{n_{r}}\left( -1\right) ^{\nu -1}\binom{n_{r}-1}{\nu
-1}\binom{n_{r}+\beta _{n_{r}}+\nu -2}{\nu }\left( 1-e^{-\alpha q}\right)
^{\nu }
\end{equation}%
where $Q_{n_{r}}=\frac{1}{2}(\frac{2}{n_{r}\alpha }-n_{r}),$ and $\beta
_{n_{r}}=1+2Q_{n_{r}}$ would leads to a target potential%
\begin{equation}
V(r)=-\frac{\sigma }{r}\text{ \ where }\sigma =\alpha (\alpha (d-1)/2+1)
\end{equation}%
with $d$-dimensional $s$ -states' eigenenergies and eigenfunctions%
\begin{equation}
E_{n_{r},0}=\frac{\alpha ^{2}}{2}Q_{n_{r}}^{2}+\mathcal{B}\text{ \ where }%
\mathcal{B}=\alpha ^{2}(4d-3)/8
\end{equation}%
\begin{equation}
R_{n_{r}}\left( r\right) =C_{n_{r}}(1+r)^{-Q_{n_{r}}-1/2}\dsum\limits_{\nu
=1}^{n_{r}}\left( -1\right) ^{\nu -1}\binom{n_{r}-1}{\nu -1}\binom{%
n_{r}+\beta _{n_{r}}+\nu -2}{\nu }\left( 1-r\right) ^{\nu }
\end{equation}

\section{Concluding Remarks}

In the point canonical transformation (PCT)\ method [23] ( an old
Liouvillean change of variables spirit [24,25]) a Schr\"{o}dinger-type
equation often mediates ( via the existence of invertible parametrization of
the real coordinates, $r\longrightarrow r\left( q\right) $, and its few
derivatives $r^{\prime }\left( q\right) ,$ $r^{\prime \prime }\left(
q\right) ,\cdots $) a transition between two different effective potentials.
In such settings, explicit correspondence (cf, e.g., Znojil and L\'{e}vai
[26]) between two bound state problems (i.e., the \emph{reference/old} and
the \emph{target/new}) is obtained. Within these Liouvillean change of
variables' spiritual lines, Alhaidari [11,12] has developed PCT-maps into 
\emph{target/new} position-dependent effective mass problems, in $d=1$ and $%
d=3$.

In this paper, a $d$-dimensional generalization of the PCT method for a
quantum particle endowed with a position-dependent mass in Schr\"{o}dinger
equation is described. Our illustrative examples include; the harmonic
oscillator, Coulomb, spiked harmonic oscillator, Kratzer-type molecular,
Morse oscillator, P\"{o}schl-Teller and Hulth\'{e}n potentials as \emph{%
reference/old} potentials to obtain exact energy eigenvalues and
eigenfunctions for \emph{target/new} potentials with different
position-dependent effective mass settings.

Finally, the applicability of the current PCT $d$-dimensional generalization
extends beyond the attendant Hermiticity settings to, feasibly, cover not
only $\mathcal{PT}$-symmetric non-Hermitian Hamiltonians but also a broader
class of $\eta $-pseudo Hermitian non-Hermitian Hamiltonians [25-28]. This
is already done in [29].


\newpage

\begin{thebibliography}{99}
\bibitem{} Puente A. and Casas M 1994 Comput. Mater Sci. \textbf{2} 441

\bibitem{} Bastard G 1988 \emph{"Wave Mechanics Applied to Semiconductor
Heterostructures" ,} Les Editions de Physique, Les Ulis

\bibitem{} Serra L I and Lipparini E 1997 Europhys. Lett. \textbf{40} 667

\bibitem{} Arias de Saaverda F, Boronat J, Polls A, and Fabrocini A 1994
Phys. Rev. \textbf{B 50} 4248

\bibitem{} Barranco M, Pi M, Gatica S.M., Hemandez E.S., and Navarro J. 1997
Phys. Rev. \textbf{B 56} 8997

\bibitem{} Puente A, Serra L I, and Casas M 1994 Z. Phys. \textbf{D 31} 283

\bibitem{} Plastino A. R., Casas M and Plastino A. 2001 Phys. Lett. \textbf{%
A281} 297 (and related references therein)

\bibitem{} van Roos O 1983 Phys. Rev. \textbf{B 27} 7547

Quesne C. 2006 Ann. Phys. \textbf{321} 1221

Tanaka T 2006 J. Phys. \textbf{A}; Math and Gen \textbf{39} 219

de Souza Dutra A \ and Almeida C. A. S.2000 Phys Lett. \textbf{A 275} 25

de Souza Dutra A 2006 J. Phys. \textbf{A}; Math and Gen \textbf{39} 203

Schmidt A 2006 Phys. Lett. \textbf{A 353} 459

\bibitem{} Dong S. H and Lozada-Cassou M. 2005 Phys. Lett. \textbf{A 337} 313

Vakarchuk I.O. 2005 J. Phys. \textbf{A}; Math and Gen \textbf{38} 4727

Cai C.Y., Ren Z.Z. and Ju G.X. 2005 Commun. Theor. Phys. \textbf{43} 1019

\bibitem{} Bagchi B, Banerjee A. Quesne C. and Tkachuk V.M. 2005 J. \ Phys. 
\textbf{A}; Math and Gen \textbf{38} 2929

Yu J. and Dong S. H. 2004 Phys. Lett. \textbf{A 325} 194

Dekar L, Chetouani L and Hammann T.F. 1998 J. Math. Phys. \textbf{39} 2551

\bibitem{} Quesne C. and Tkachuk V.M. 2004 J. \ Phys. \textbf{A}; Math and
Gen \textbf{37} 4267

Jiang L. Yi L.-Z., and Jia C.-S. 2005 Phys. Lett. \textbf{A 345} 279

Alhaidari A.D. 2003 Int. J. Theor. Phys. \textbf{42} 2999

\bibitem{} Alhaidari A.D. 2002 Phys. Rev. \textbf{A 66} 042116

\bibitem{} De R., Dutt R.and Sukhatme U 1992 J. \ Phys. \textbf{A}; Math and
Gen \textbf{25} L843

\bibitem{} Junker G 1990 J. \ Phys. \textbf{A}; Math and Gen \textbf{23} L881

\bibitem{} Znojil M and L\'{e}vai G. 2001 J. Math. Phys. \textbf{42} 1996

Lucht M W and Jarvis P D 1993 Phys. Rev. \textbf{A 47} 817

Ushveridze A G 1994 \emph{Quasi-exactly Solvable Models in Quantum Mechanics,%
} IOP, Bristol, England.

\bibitem{} Herschbach D R et al 1993 \emph{"Dimensional Scaling in Chemical
Physics" }( Kluwer Academic Publishers, Dordrecht, Netherlands.)

Herschbach D R 1986 J. Chem. Phys. \textbf{84} 838

Taseli H 1996 J. Math. Chem. \textbf{20} 235

\bibitem{} Mustafa O and Znojil M 2002 J. Phys.A; Math and Gen. \textbf{35}
8929

Mustafa \thinspace O and Odeh M 1999 J. Phys. A; Math and Gen. \textbf{32}
6653

\bibitem{} Gang C 2004 Phys. Lett. \textbf{A 329} 22

Nieto M. M. 1979 Am. J. Phys. \textbf{47} 1067

\bibitem{} Chen G 2004 Phys. Lett. \textbf{A 326} 55

Sun H 2005 Phys. Lett. \textbf{A 338} 309

\bibitem{} Mustafa \thinspace O and Odeh M 2000 J. Phys. A; Math and Gen. 
\textbf{33} 5207

Hall R L and Saad N 2000 J. Phys. A; Math and Gen \textbf{33} 569

Hall R L and Saad N 1999 J. Phys. A; Math and Gen \textbf{32} 133

Ciftci H, Hall R L and Saad N 2003 J. Phys. A; Math and Gen. \textbf{36}
11816

Hall R L, Katatbeh Q D and Saad N 2004 J. Phys. A; Math and Gen. \textbf{37}
11629

\bibitem{} Salem L D and Montemayor R 1993 Phys. Rev. \textbf{A 47} 105

Fl\"{u}gge S 1974, \emph{Practical Quantum Mechanics,} Springer, Berlin,

\bibitem{} Lam C S and Varshni Y P 1971 Phys. Rev. \textbf{A 4} 1875

\bibitem{} Pak N K and S\"{o}kmen I 1984 Phys. Lett. \textbf{A 103} 298

\bibitem{} Liouville J 1837 J. Math. Pure Appl. \textbf{1} 16

\bibitem{} Znojil M and L\'{e}vai G 2001 J. Math. Phys. \textbf{42} 1996

Znojil M 2000 "$\mathcal{PT}$-symmetric form of the Hulth\'{e}n potential"
(arXiv: math-ph/0002017)

\bibitem{} Znojil M and L\'{e}vai G 2000 Phys. Lett. \textbf{A 271} 327

\bibitem{} Mostafazadeh A 2002 J. Math. Phys. \textbf{43} 205

Mostafazadeh A 2002 J. Math. Phys. \textbf{43} 3944

Mostafazadeh A 2003 J. Math. Phys. \textbf{44} 974

Ahmed Z 2001 Phys. Lett. \textbf{A 2290} 19

\bibitem{} Znojil M 2002 "Pseudo-Hermiian version of the charged harmonic
oscillator and its "forgotten" exact solution" (arXiv: quant-ph/0206085

Sinha A and Roy P 2003 "Generalization of exactly solvable non-Hermitian
potentials with real energies" (arXiv: quant-ph/0312089)

Mustafa O and Mazharimousavi S. H. 2006 "$\eta $-Pseudo-Hermiticity
generators: radially symmetric Hamiltonians" (arXiv: hep-th/0601017)

\bibitem{} Mustafa O and Mazharimousavi S. H. 2006 Czech. J. Phys. (in
press) (arXiv: quant-ph/0603272)
\end{thebibliography}
\end{document}